\documentclass[sigconf,screen,authorversion]{acmart}

\usepackage{xcolor}
\usepackage{algorithm}
\usepackage{algorithmic}
\usepackage{amsmath} % for mathematical symbols and environments
\usepackage{siunitx} % for aligning numerical data in table with decimal point 
\usepackage{bm} % for bolding changes in Equation 1
\usepackage{balance}

\definecolor{lightblue}{rgb}{0.8, 0.9, 1}

\AtBeginDocument{%
  \providecommand\BibTeX{{%
    \normalfont B\kern-0.5em{\scshape i\kern-0.25em b}\kern-0.8em\TeX}}}

\setcopyright{acmlicensed}
\acmDOI{10.1145/3650212.3680382}
\acmYear{2024}
\copyrightyear{2024}
\acmISBN{979-8-4007-0612-7/24/09}
\acmConference[ISSTA '24]{Proceedings of the 33rd ACM SIGSOFT International Symposium on Software Testing and Analysis}{September 16--20, 2024}{Vienna, Austria}
\acmBooktitle{Proceedings of the 33rd ACM SIGSOFT International Symposium on Software Testing and Analysis (ISSTA '24), September 16--20, 2024, Vienna, Austria}
\acmSubmissionID{issta24main-p1316-p}
\received{2024-04-12}
\received[accepted]{2024-07-03}

\begin{document}

\title{Policy Testing with MDPFuzz (Replicability Study)}

\author{Quentin Mazouni}
\orcid{0009-0003-3519-5514}
\affiliation{%
  \institution{Simula Research Laboratory}
  \city{Oslo}
  \country{Norway}
}
\email{quentin@simula.no}

\author{Helge Spieker}
\orcid{0000-0003-2494-4279}
\affiliation{%
  \institution{Simula Research Laboratory}
  \city{Oslo}
  \country{Norway}
}
\email{helge@simula.no}

\author{Arnaud Gotlieb}
\orcid{0000-0002-8980-7585}
\affiliation{%
  \institution{Simula Research Laboratory}
  \city{Oslo}
  \country{Norway}
}
\email{arnaud@simula.no}

\author{Mathieu Acher}
\orcid{0000-0003-1483-3858}
\affiliation{%
  \institution{Univ Rennes, CNRS, Inria, IRISA
  \\Institut Universitaire de France (IUF)
  }
  \city{UMR 6074, F-35000 Rennes}
  \country{France}
}
\email{mathieu.acher@irisa.fr}

%%
%% The abstract is a short summary of the work to be presented in the
%% article.
\begin{abstract}

In recent years, following tremendous achievements in Reinforcement Learning, a great deal of interest has been devoted to ML models for sequential decision-making.
Together with these scientific breakthroughs/advances, research has been conducted to develop automated functional testing methods for finding faults in black-box Markov decision processes.
Pang et al. (ISSTA 2022) presented a black-box fuzz testing framework called MDPFuzz.
The method consists of a fuzzer whose main feature is to use {\it Gaussian Mixture Models} (GMMs) to compute coverage of the test inputs as the likelihood to have already observed their results.
This guidance through coverage evaluation aims at favoring novelty during testing and fault discovery in the decision model.

Pang et al. evaluated their work with four use cases, by comparing the number of failures found after twelve-hour testing campaigns with or without the guidance of the GMMs (ablation study). 
In this paper, we verify some of the key findings of the original paper and explore the limits of MDPFuzz through reproduction and replication.
We re-implemented the proposed methodology and evaluated our replication in a large-scale study that extends the original four use cases with three new ones.
Furthermore, we compare MDPFuzz and its ablated counterpart with a random testing baseline.
We also assess the effectiveness of coverage guidance for different parameters, something that has not been done in the original evaluation. 
Despite this parameter analysis and unlike Pang et al.'s original conclusions, we find that in most cases, the aforementioned ablated Fuzzer outperforms MDPFuzz, and conclude that the coverage model proposed does not lead to finding more faults.

\end{abstract}

%%
%% The code below is generated by the tool at http://dl.acm.org/ccs.cfm.
%% Please copy and paste the code instead of the example below.
%%


%%
%% Keywords. The author(s) should pick words that accurately describe
%% the work being presented. Separate the keywords with commas.

%% A "teaser" image appears between the author and affiliation
%% information and the body of the document, and typically spans the
%% page.
% \begin{teaserfigure}
%   \includegraphics[width=\textwidth]{sampleteaser}
%   \caption{Seattle Mariners at Spring Training, 2010.}
%   \Description{Enjoying the baseball game from the third-base
%   seats. Ichiro Suzuki preparing to bat.}
%   \label{fig:teaser}
% \end{teaserfigure}

%%
%% This command processes the author and affiliation and title
%% information and builds the first part of the formatted document.
\maketitle

\section{Introduction}

In recent years, the combination of Machine Learning (ML) with neural networks has been effectively employed to solve complex sequential decision-making tasks of diverse nature, such as gaming~\cite{doi:10.1126/science.aar6404}, planning~\cite{DBLP:journals/corr/abs-1908-01362, DBLP:journals/corr/abs-2007-06702} or system control~\cite{dietterich2000hierarchical, 8675643}.
Likewise, dedicated testing techniques have been proposed to enable the safe use and deployment of these models.
Despite the wide variety in the current approaches, such as search-based testing~\cite{https://doi.org/10.48550/arxiv.2205.04887}, metamorphic testing~\cite{10.1145/3533767.3534392, Eisenhut2023AutomaticMT} and fuzz testing~\cite{Steinmetz_Fišer_Eniser_Ferber_Gros_Heim_Höller_Schuler_Wüstholz_Christakis_Hoffmann_2022, 10.1145/3533767.3534388}, few works are comparing them, exploring their limitations or verifying their findings. 
In particular, as time of writing no replicability study has been conducted. 

In 2022,~\citeauthor{10.1145/3533767.3534388}~\cite{10.1145/3533767.3534388} presented a black-box fuzz framework called MDPFuzz.
We decided to replicate this precise framework for several reasons.
First, MDPFuzz is generally applicable and shown to be effective at finding functional faults in decision-making policies.
Second, the method has attracted strong interest in the software testing community.
For instance, several works have built upon it~\cite{curefuzz, seqdivfuzz}. 
Similarly,~\citeauthor{model_based_fuzzer}~\cite{model_based_fuzzer} and~\citeauthor{ast2024}~\cite{ast2024} consider MDPFuzz as a state-of-the-art policy testing technique to evaluate their approach.

However, we identified severe bugs in the original implementation as well as significant differences compared to the algorithm of the method (detailed later).
We contacted the authors to discuss these discrepancies (during an email conversation~\cite{emailconversation}).
Finally, we worried about its functional complexity, something critical for efficient testing.
% "functional complexity" "implementation hardness".
Indeed, this framework guides the testing process with coverage estimations based on Gaussian Mixture Models (GMMs)~\cite{10.2307/2982840} which are compute-heavy tasks.
% Its main features are its use of Gaussian Mixture Models (GMMs)~\cite{10.2307/2982840} and an algorithm to update them at a constant cost.

Precisely, the idea of MDPFuzz is to guide a fuzzing process~\cite{10.1145/96267.96279} that generates and mutates input tests.
% The guidance consists of two mechanisms: (1) keeping inputs that reveals weakness in the model under test (i.e. robustness) and (2) keeping inputs whose test result is deemed uncovered (i.e., coverage-based guidance).
Testing is guided by maintaining the pool of inputs with (1) ones that reveal a weakness in the model under test (i.e. robustness) and (2) ones whose test result is deemed uncovered (i.e. coverage-based guidance).
~\citeauthor{10.1145/3533767.3534388} propose to compute input coverage over the state sequence resulting from a test case with two GMMs, which requires the computation of $1+2\text{ }|M|$ probability densities ($M$ being the length of the sequence).
The additional coverage guidance aims at exercising the model under test in novel ways such that faults are more likely to be found.
\citeauthor{10.1145/3533767.3534388} evaluated their work with four use cases on fault detection, GMMs guidance efficiency, fault analysis and policy improvement (retraining). 
In particular, they show that MDPFuzz finds more faults than an ablated version we refer to as Fuzzer.
This framework is only guided by robustness, i.e., it maintains its pool with mechanism (1) and, consequently, does not have GMMs and does not compute coverage.

In this paper, we investigate the ability of MDPFuzz to find faults in decision-making models. 
Given the issues and algorithmic discrepancies found in the original implementation, we decided to conduct a two-step methodology. 
In the first step (later called ``reproduction study''), we patch the code and use both the unfixed and fixed versions in an attempt to reproduce the original findings and determine whether the bugs were present when~\citeauthor{10.1145/3533767.3534388} evaluated their method.
In a second step (later called ``replication study''), we re-implement the fuzzers, i.e., we implement the algorithm described in the original paper and extend their evaluation.
Specifically, We add three new use cases and include a random testing baseline.
Furthermore, we perform a parameter analysis of MDPFuzz -- which is not available in the original publication -- to (1) assess its sensitivity, (2) mitigate potential biases in the previous results and (3) possibly find out an optimal configuration for fault discovery.

% In this paper, we replicate the proposal of~\citeauthor{10.1145/3533767.3534388}~\cite{10.1145/3533767.3534388} and extend the original evaluation to explore the limitations of MDPFuzz.
% We focus on the use of MDPFuzz to find faults in the decision models and further investigate the need for coverage guidance (ablation study).
% To that end, we conduct a two-step, reproduction-replication study, in which we use the original implementation before re-implementing the proposal. 
% We use our replicate to compare the two versions of the fuzzers (with and without coverage guidance) to random testing -- which is not available in the original publication -- in an extended evaluation with three new use-cases.
% Eventually, we perform a parameter analysis of the coverage model to find out optimal values and mitigate potential biases in our results.

% Our initial intention was to first reproduce their results (thanks to the source code provided by the authors), and then to replicate their proposal to extend its evaluation.
% However, when we reviewed the code of their experiments we noticed a lot of significant changes compared to the framework proposed in the paper.
% Moreover, we had difficulties to reproduce some of the experiments themselves.
% Therefore, we decided to implement their paper on our own to check whether their approach is actually working on the original use-cases in the first place.
% Unfortunately, despite our best efforts, our experiments systematically showed deceptive results.

In summary, the contributions of this paper are:
\begin{itemize}
    \item \textbf{Reproduction} We report on the issues and difficulties in reproducing the results with the original implementation. 
    We describe our best efforts to understand and mitigate them and identify the underlying threats to the validity of the original evaluation.
    % We describe the differences between the implementation of MDPFuzz and its specification defined in the original publication, the (unintentional) bugs and the missing parts
    We patch the code and present the results obtained with both the unfixed and fixed versions.
    \item \textbf{Replication} We re-implement MDPFuzz and its coverage-free version. Our results validate the fault discovery ability of MPDFuzz in the tested models, but unlike the original evaluation, we find that the ablated version outperforms its coverage-guided counterpart.
    \item \textbf{Comparison} We compare our replication of the two fuzzers with a random testing baseline in an extended study of seven use cases.
    % \item \textbf{Parameter Evaluation} The coverage model embedded by MDPFuzz is parameterized by three parameters, yet their values are not straightforward to define nor were they investigate in the original evaluation. 
    % We thus investigate their impact in terms of fault detection.
    \item \textbf{Parameter Evaluation} The coverage model integrated by MDPFuzz is parameterized by three parameters, but defining their values is not straightforward, and they were not investigated in the original evaluation. Therefore, we explore their impact on fault detection.
    \item \textbf{Valuable Artifacts} Following the best practices of replication studies, our replicate is freely available.
    In particular, we provide the core components of MDPFuzz as reusable modules to enable convenient future work. 
    % Our replicate can be easily installed through the package manager \textsc{PIP}.
\end{itemize}

% The rest of this paper is organized as follows.
% Section~\ref{sec:background} introduces policy testing and MDPFuzz~\cite{10.1145/3533767.3534388}.
% We explain the benefits of replicating the work of~\citeauthor{10.1145/3533767.3534388}~\cite{10.1145/3533767.3534388} and detail our motivations in Section~\ref{sec:motivation}.
% Section~\ref{sec:reproduction} describes our attempt to reproduce the original work with the implementation publicly released.
% In Section~\ref{sec:replication}, we then present how we eventually replicate from scratch the proposal and the results of our exhausting evaluation.
% We address the limitation of our investigation in Section~\ref{sec:threats} and conclude in Section~\ref{sec:conclusion}. 

\section{Background}
\label{sec:background}

In this section, we introduce policy testing for sequential decision-making and MDPFuzz.

\subsection{Sequential Decision-Making}

Roughly speaking, sequential decision-making describes any task that can be solved in a step-by-step manner by one or several models~\cite{FrankishRamsey2014}.
In this work, we are interested in testing these decision-making entities.
Precisely, a test case consists of setting the environment of the decision task to an initial state and letting the model under test (or agent) interact with the former.
This step-wise, observation-decision-action process eventually leads to a final state, which is passed to the test oracle to detect a failure.
Decision-making problems are formally defined as Markov Decision Processes (MDPs), often represented as 4-tuples $\langle S,A,R,T\rangle$:
\begin{itemize}
    \setlength{\parskip}{2pt}
    \item $S$ is a set of states. They correspond to what the agent observes at every step of an execution.
    \item $A$ is the set of possible actions, which can be either continuous or discrete. They are what the agent returns to the environment. 
    \item $R:S\times A\mapsto\mathbb{R}$ is the reward function. Rewards are the feedback received by the agent after every action. 
    \item $T:S\times A\times S\mapsto[0,1]$ is the transition function, which is a probability distribution over $S$ and $A$. It depicts which state the MDP will transit to for a given action. 
    The function is not known by the agent and describes the environment's dynamic.
\end{itemize}
Solutions to MDPs are mapping functions $\pi:S\times A\mapsto[0,1]$ called policies, which choose the action for a given state.
The most commonly used approach to train policies is Reinforcement Learning (RL), a sub-field of Machine Learning which consists of learning from the rewards/penalties returned by the MDP~\cite{sutton2018reinforcement}.
Precisely, RL learns an optimal, or near-optimal, policy that maximises the total expected discounted cumulative reward $R_t=\sum_{t>0} \gamma^{t-1}r_t$, where $\gamma\in[0,1)$ is the discount factor.
The latter controls how $\pi$ considers future rewards: a small value emphasizes short-term gains, whereas values close to 1 lead the agent to focus on maximising long-term rewards.

In this work, we consider black-box policy testing, meaning that, during a test case execution, we can only observe the interaction between MDP and agent, but have no access to any of their internals. 
MDPFuzz further assumes that the executions are limited to $M$ steps, and targets \textit{deterministic} decision models.
As such, given the model under test $\pi:S\mapsto A$, we denote the execution of input $i\in S$ as $MDP(i,\pi)$.
The result is composed of the cumulative reward $r$ and the sequence of states $\{s_t\}_{t\in 0..M\text{-}1}$.

\subsection{MDPFuzz}\label{back:mdpfuzz}

\subsubsection{Preliminaries}

MDPFuzz is based on fuzzing, which here means that the inputs (or initial states) used to initialize the MDP are first randomly generated and then mutated.
To that end, the method records the inputs tested in a pool along with two measurements: sensitivity (also called ``energy'') and their coverage (or ``density'').

\paragraph{Sensitivity} 
Sensitivity estimates the robustness of $\pi$ against MDPFuzz's mutations on the states. 
The sensitivity of the state $s$ is defined as the absolute reward difference between $MDP(s,\pi)$ and $MDP(s+\Delta s,\pi)$, where $\Delta s$ is a small random perturbation~\cite{10.1145/3533767.3534388}.
The higher the sensitivity of a state, the more diverse the results of its mutations.
The sensitivities are used to bias the input selection during fuzzing, which we discuss below.  

\paragraph{Coverage}
This measure aims at fostering the novelty of the state sequences collected after test executions.
It aims at favoring the discovery of new behaviors by keeping low-covered inputs back in the pool.
Precisely, the coverage $d_s$ of an initial state $s$ is defined as the following joint probability density:
\begin{equation}\label{eq:coverage}
    d_s=p(s_0)\times\prod^{M-2}_{t=0}\frac{p(s_t,s_{t+1})}{p(s_t)}
\end{equation}
where $\{s_t\}_{t\in 0..M\text{-}1}$ is the state sequence returned by $MDP(s,\pi)$.
\citeauthor{10.1145/3533767.3534388}~\cite{10.1145/3533767.3534388} propose to estimate the probability densities $p(.)$ with two Gaussian Mixture Models (GMMs)~\cite{10.2307/2982840}, dealing with single states $s_t$ and concatenated states $s_t,s_{t+1}$, respectively. 
% The parameters of GMMs are commonly approximated with Expectation Maximization~\cite{https://doi.org/10.1111/j.2517-6161.1977.tb01600.x}.
% A key feature of MDPFuzz is \textit{DynEM}, an algorithm that avoids re-computing GMMs' parameters after every test case by dynamically updating the current ones.
A common approach to approximate the parameters of GMMs is Expectation Maximization~\cite{https://doi.org/10.1111/j.2517-6161.1977.tb01600.x}; however, this method requires their re-computation after every test case execution.
A key feature of MDPFuzz is \textit{DynEM}, an algorithm that dynamically updates the current parameters at a constant cost.
For further details, interested readers can refer to the original work~\cite{10.1145/3533767.3534388}. 

% \hfill
\subsubsection{Method}

Algorithm~\ref{alg:mdpfuzz} illustrates the high-level working of MDPFuzz.
The method inputs a number of initial states $N$, the model under test $\pi$ as well as three parameters: the threshold to update the GMMs (with DynEM) $\tau$, the update weight $\gamma$ and $K$, the number of components of the GMMs. 

\paragraph{Initialization} 
This stage is described at lines~\ref{start_init}-\ref{end_init}, and aims at feeding the pool with inputs and their sensitivities and coverage. 
It begins by initializing the parameters of the GMMs and the dynamic update algorithm DynEM.
Then, $N$ initial states are randomly sampled in the observation space of the MDP.
MDPFuzz iterates the states to compute and record their respective sensitivity and coverage.
When the latter is lower than $\tau$, the parameters of the GMMs are updated with DynEM.

\paragraph{Fuzzing}
This second and main stage of the framework starts at line~\ref{fuzzing}.
It consists of a loop in which three key steps are performed.
First, an input $s$ is selected from the pool and mutated (lines~\ref{select}-\ref{mutate}).
As mentioned above, the selection is biased with respect to the sensitivity of the states.
Precisely, the probability of selecting the input $s_i$ of sensitivity $e_i$ is $\frac{e_i}{\sum^{n-1}_{i=0}e_i}$, where $n$ is the current size of the pool.
Then, $\pi$ is tested with the mutated input (line~\ref{mut_exec}).
The latter is saved in the solution set in case of a failure.
In the third step, the mutated input is added to the pool if (1) it induces a lower reward than its original counterpart or (2) its coverage is lower than $\tau$. 
As in the previous stage, in such a case the parameters of the GMMs are updated with DynEM.

\definecolor{blue}{RGB}{0, 20, 220}
\definecolor{cc}{RGB}{125, 125, 125}

\begin{algorithm}
\renewcommand{\algorithmiccomment}[1]{\textit{\textcolor{cc}{#1}}}
\newcommand{\LONGCOMMENT}[1]{\textit{\textcolor{cc}{$\triangleright$ \: #1 \hfill $\triangleleft$}}}
\renewcommand{\algorithmicrequire}{\textbf{Input:}}
\renewcommand{\algorithmicensure}{\textbf{Output:}}
\caption{MDPFuzz}
\label{alg:mdpfuzz}
\begin{algorithmic}[1]
\REQUIRE $N$: initial size of the pool, $\pi$: model under test, $\tau$: coverage threshold, $\gamma$: update weight, $K$: number of components for the 2 GMMs
\ENSURE $R$: crash-revealing initial states
\STATE $Pool \text{, } R \gets \emptyset \text{, } \emptyset$\label{start_init}
\STATE \LONGCOMMENT{initialize GMMs' and DynEM's parameters}
\STATE $GMM^s, GMM^c \gets \textcolor{blue}{\textbf{init\_DynEM}}(K,\gamma)$ 
\STATE \LONGCOMMENT{sample N initial states to feed the pool}
\FOR{$i=1$ \TO $N$}\label{alg:init_sampling}
    \STATE $s_i \gets \textcolor{blue}{\textbf{sample\_initial\_state}}()$
    % \STATE \LONGCOMMENT{compute the sensitivity of $S_i$}
    \STATE $e_i \gets \textcolor{blue}{\textbf{sensitivity}}(s_i)$
    % \STATE \LONGCOMMENT{execute $S_i$ and collect the reward and state sequence}
    \STATE $r_i \text{, } \{s_t\}_{t\in 0..M\text{-}1} \gets \textcolor{blue}{\textbf{MDP}}(s_i,\pi)$
    \STATE $d_i \gets \textcolor{blue}{\textbf{coverage}}(\{s_t\}_{t\in 0..M\text{-}1}, GMM^s, GMM^c)$
    \STATE \LONGCOMMENT{update GMMs' parameters if low coverage}
    \IF{$d_i < \tau$}
        \STATE $GMM^s, GMM^c \gets \textcolor{blue}{\textbf{DynEM}}(\{s_t\}_{t\in 0..M\text{-}1}, GMM^s, GMM^c)$
    \ENDIF
    \STATE $Pool.add(s_i, r_i, e_i, d_i)$ \COMMENT{\hspace{1em} $\triangleright$ feed the pool}
\ENDFOR\label{end_init}
\STATE \LONGCOMMENT{fuzz until test budget is consumed}
\WHILE{$\text{test budget}$}\label{fuzzing}
    \STATE \LONGCOMMENT{sensitivity-biased random selection}
    % \STATE $S, r \gets \textcolor{blue}{\textbf{SELECT}}(Pool)$
    \STATE $s \text{, } r \gets Pool.select()$\label{select}
    \STATE \LONGCOMMENT{mutate and execute the state}
    \STATE $s' \gets \textcolor{blue}{\textbf{mutate}}(s)$\label{mutate}
    \STATE $r' \text{, } \{s_t\}_{t\in 0..M\text{-}1} \gets \textcolor{blue}{\textbf{MDP}}(s',\pi)$\label{mut_exec}
    % \STATE \LONGCOMMENT{compute the coverage}  
    % \STATE $f' \gets \textcolor{blue}{\textbf{SEQ\_FRESH}}(\{S_t\}_{t\in[M-1]}, Params^s, Params^c, \tau)$  
    \STATE $d' \gets \textcolor{blue}{\textbf{coverage}}(\{s_t\}_{t\in 0..M\text{-}1}, GMM^s, GMM^c)$
    \STATE \LONGCOMMENT{detect and record fault}  
    \IF{$\textcolor{blue}{\textbf{crash}}(\{s_t\}_{t\in 0..M\text{-}1})$}
        \STATE $R.add(s')$
    \STATE \LONGCOMMENT{maintain the pool if lower reward or low coverage}  
    \ELSIF{$r' < r\textbf{ or }d' < \tau$}\label{alg:cond} 
        \STATE $e' \gets \textcolor{blue}{\textbf{sensitivity}}(s')$
        \STATE $Pool.add(s', r', e', d')$
        \STATE \LONGCOMMENT{dynamic update of GMMs' parameters}
        \STATE $GMM^s, GMM^c \gets \textcolor{blue}{\textbf{DynEM}}(\{s_t\}_{t\in 0..M\text{-}1}, GMM^s, GMM^c)$
    \ENDIF
\ENDWHILE
\RETURN $R$
\end{algorithmic}
\end{algorithm}

In their work,~\citeauthor{10.1145/3533767.3534388} compare MDPFuzz to what we refer as the ``Fuzzer'', which is an ablated version that does not compute coverage. 
As such, this framework only measures sensitivities and maintains its pool with (1), i.e., mutated inputs that provoke lower rewarded executions than their original counterparts. 
Note that by being coverage-free, Fuzzer is not parameterized.

\section{Research Questions}

Our primary goal is to verify the ability of MDPFuzz to find functional faults in the models under test, both with the original implementation (provided by~\citeauthor{10.1145/3533767.3534388}) and with our re-implementation.
As detailed in introduction, we followed this two-step methodology after experiencing issues with the original code.

Our secondary goal is to investigate the effects of the parameters $K$, $\tau$ and $\gamma$ on MDPFuzz's performance.
In short, this work answers the following research questions:
\begin{enumerate}
    \item[\textbf{RQ1}] Reproduction study: can we reproduce the results with the original implementation?
    \item[\textbf{RQ2}] Replication study: can our replicate discover faults in the original use cases, in other use cases? How does it compare to Random Testing and Fuzzer (also re-implemented)?
    \item[\textbf{RQ3}] Parameter analysis: is MDPFuzz sensitive to its parameters? Can we find optimal configurations for fault detection?
\end{enumerate}

Answering the two first research questions will let us conclude on the general fault discovery ability of MDPFuzz and Fuzzer.
In particular, the replicate, i.e., RQ2, will compare these techniques against Random Testing, which has not been done in the original evaluation. 
% The last question investigates whether we can find optimal parameters for the coverage model that would drive testing towards more faults. 
The last question examines the sensitivity of MDPFuzz to its parameters, and how they affect coverage guidance for fault detection. 
Besides, it will let us check that the original evaluation as well as our reproduction study (RQ1, in Section~\ref{sec:reproduction}) have not been biased by weak parameters.

From now on, we distinguish the results of the reproduction study (using the original implementation) from the replication using our re-implementation, (cf. Section~\ref{sec:replication}) by applying the suffix ``-O'' and ``-R'', respectively. 

\section{Reproduction Study}\label{sec:reproduction}

In the following, we explain our methodology to attempt to reproduce the results of~\citeauthor{10.1145/3533767.3534388} with the original implementation\footnote{\url{https://github.com/Qi-Pang/MDPFuzz}}.
Indeed, we encountered difficulties in using the implementation, and we had to add and change some parts of the code.

The first subsection documents the two most significant differences we found between the implementation and the original specification of MDPFuzz.
The second presents the findings of our thorough code review where we cover all the bugs identified as well as the missing or inconsistent elements that prevented us from running the experiments.
The third subsection details the subsequent additions and modifications we had to perform to successfully execute the experiments, comments on our results and their deviations from the original results.

\subsection{Algorithmic Differences}\label{subsec:changes}

The two most significant discrepancies compared to the specification of MDPFuzz defined in the original publication concern coverage computation and parameter update (DynEM).
In the following, we present those changes and the authors' explanation.

\paragraph{Coverage computation} The authors consider the \textit{probability densities} (computed by the GMMs) as \textit{probabilities}.
% As such, every quotient in Equation~\ref{eq:coverage} is bounded between 0 and 1, turning thus the definition into:
Indeed, instead of implementing the definition of Equation (1), the authors have implemented the following definition:
$$d_s=p(s_0)\times\prod^{M-2}_{t=0}\bm{min(1,}\frac{p(s_t,s_{t+1})}{p(s_t)}\bm{)}$$
% We don't understand that decision since the values calculated by the GMMs are undoubtedly densities, which are not meant to belong to $[0, 1]$.
We think that the introduction of the $min$ function corresponds to a wanted simplification (i.e., real variables in $[0,1]$).

\paragraph{GMMs' parameters update} DynEM updates the parameters of the GMMs with all the states of a given sequence~\cite{10.1145/3533767.3534388}. 
However, we found that the mixture models for the single and concatenated states are updated with the first state and 10\% of the states, respectively.
The authors argued that it is a good trade-off for efficiency and accuracy since the GMMs aim at approximating correlations between the states~\cite{emailconversation}. 
They further detailed that the mixture models are not powerful enough to estimate long sequences, which could lead to overestimation.

\subsection{Code Review}\label{subsec:review}

In the following, we present the unintentional bugs we found in the original code and the missing or inconsistent elements that made us unable to perform the original evaluation without addressing them (i.e., changing the code).

\paragraph{Bugs in Coop-Navi} We found two severe bugs in the implementation of the Coop-Navi case study.
First, the pool is fed with \textit{references} of the initial states instead of copies. 
Because the state of the MDP evolves during test execution, the pool ends up being fed with \textit{final} states, i.e., a failure or solved state.
In any case, mutating such states is likely to produce very short executions.
The second issue concerns the computation of inputs' sensitivity during the sampling phase (described in Subsection~\ref{back:mdpfuzz}).
We found that, even though the perturbed inputs are \textit{computed}, they are not then \textit{used} to reset the MDP.
In other words, if the first flaw were fixed, then the sensitivities would always be null (same input re-executed). 

\paragraph{Flawed mutation in Bipedal Walker} The instructions that ensure that the mutated and input states are different are put in comments (i.e., they are not executed).
We empirically computed the probability of null mutation as 20\%.
% As with the previous suspicion, we cannot tell if they were already commented when the experiments were conducted.

\paragraph{Missing Fuzzer baselines} The Fuzzer baseline is only implemented for the Bipedal Walker case study, in which coverages are still computed but not used, thus artificially decreasing its efficiency.
Such biased implementation follows what~\citeauthor{10.1145/3533767.3534388} suggested to do when we asked why the aforementioned baselines were missing: to simply change the condition to maintain the pool during fuzzing (i.e., line~\ref{alg:cond} in Algorithm~\ref{alg:mdpfuzz}).

\paragraph{Different initialization requirements} The initialization procedure implemented requires a number of inputs to sample, as specified in Algorithm~\ref{alg:mdpfuzz}, line~\ref{alg:init_sampling}.
However, the one defined in the evaluation section of the original work consists of 2 hours of sampling.
We asked the authors for clarification and they confirmed that the sampling was performed for 2 hours.
They didn't explain to us why it is not the case in the code.

\paragraph{Different Hyperparameters} While the original publication does not mention the parameters of MDPFuzz evaluated, in a preprint\footnote{Available here: \url{https://arxiv.org/abs/2112.02807v3}.}  their values are $K=10$ and $\tau=\gamma=0.01$.
% Surprisingly though, the code implements different values (we don't detail due to space limitations).
Surprisingly though, the code implements different values (summarized in Table~\ref{tab:params}).
As before, we sought clarification from the authors and they confirmed that they used the values $K=10$, $\tau=0.01$ and $\gamma=0.01$.
They added that the code has been later modified for testing/debugging purposes.

\begin{table}[h]
\caption{Parameters of MDPFuzz found in original implementation for each use case. 
The last row shows the values confirmed by the authors.}
\label{tab:params}
\Description[Parameters of MDPFuzz found in original implementation for each use case.]{}{}
\begin{tabular}{lrrr}
\toprule
% Use Case       & $K$  & $\tau$  & $\gamma$ \\\midrule
Use Case       & $K$  & $\tau$  & $\gamma$ \\ \midrule % \cmidrule(lr){2-4}
ACAS Xu        & 5  & 0.1  & 0.1   \\
Bipedal Walker & 10 & 0.02 & 0.1   \\
CARLA          & 10 & 0.1  & 0.1   \\
Coop Navi      & 1  & 0.01 & 0.1   \\ \midrule %\cmidrule(lr){1-1}
Paper~\cite{10.1145/3533767.3534388}         & 10 & 0.01 & 0.01 \\\bottomrule
\end{tabular}
% \vspace{0.2cm}
\end{table}

Besides, we note that the underlying MDPs of the Bipedal Walker and CARLA use cases include randomness and are thus stochastic.
To that regard, the authors could have decided to reset the seed of the random generator for Bipedal Walker, while -- as far as we know -- there is no way to entirely control the randomness of the CARLA simulator\footnote{In particular, the next direction of the vehicles when they enter an intersection is randomly chosen and cannot be controlled.}.
If considering stochastic $MDP(.)$ is not a bug, it might lead to unstable results and disables replicable test case execution.
Hence, in our replication study (RQ2, in Section~\ref{sec:replication}), we fix the random parameters of the underlying MDP (i.e., given an input $i$, $MDP(i,\pi)$ returns the same results).

\subsection{Reproduction Results}\label{subsec:repro}

\paragraph{Implemented Changes}
To account for the authors' answers and the bugs discovered, we forked the original code and made the following changes.
% First, we over-headed the code with our own logging methods, since no script to analyze the initial logs is provided.
First, we added logging methods, since no script to analyze the initial logs is provided.
Second, we set back the GMM's and DynEM's parameters to the values confirmed by the authors~\cite{emailconversation}, namely $K=10$, $\tau=0.01$, and $\gamma=0.01$.
Similarly, we changed the initialization functions such that the fuzzers sample for 2 hours.
Third, we added the Fuzzer for all the use cases, by copy-pasting the exact code of MDPFuzz and putting in comments all the instructions related to coverage guidance and computation.
Eventually, we fixed the three bugs detailed in Subsection~\ref{subsec:review}. 
As a last note, even though we expected the original implementation to contain all the material to reproduce~\citeauthor{10.1145/3533767.3534388}' experiments, we emphasize that the artifacts provided are labelled as ``available'' (and not, for instance, as ``reproducible'') and note that~\citeauthor{10.1145/3533767.3534388} kindly answered our interrogations.

\paragraph{Experimental Design}
To assess how the bugs threaten the validity of the original evaluation, we consider both the unmodified and the modified versions of MDPFuzz-O and Fuzzer-O in the reproduction results.
Precisely, for Bipedal Walker, we evaluate the untouched mutation function and the fixed one (later referred as ``Mutation''), for which null mutation is not possible anymore. 
For the Coop Navi case study, we compare three versions: original, ``Sampling'' (where the pools are fed with the initial states) and ``Fixed'' (the two bugs are addressed).
We did not study the fourth case, which would isolate the second issue, because of fatal errors.
Indeed, not perturbing initial states systematically implies null sensitivity; this corner case is not handled by the implementation.

We follow the experimental procedure defined in the original paper and account for the randomness effect by running each execution three times.
All the material of the experiments is available online\footnote{\url{https://github.com/QuentinMaz/MDPFuzz_Replicability_Study_Artifact/tree/master/reproduction}}.
We ran the experiments on a Linux machine (Ubuntu 22.04.3 LTS) equipped with an AMD Ryzen 9 3950X 16-Core processor and 32GB of RAM.
Due to long execution times, we doubled the sampling and fuzzing times for the CARLA use case compared to the original paper.

\begin{figure*}[t]
    \centering
    \includegraphics[width=\textwidth]{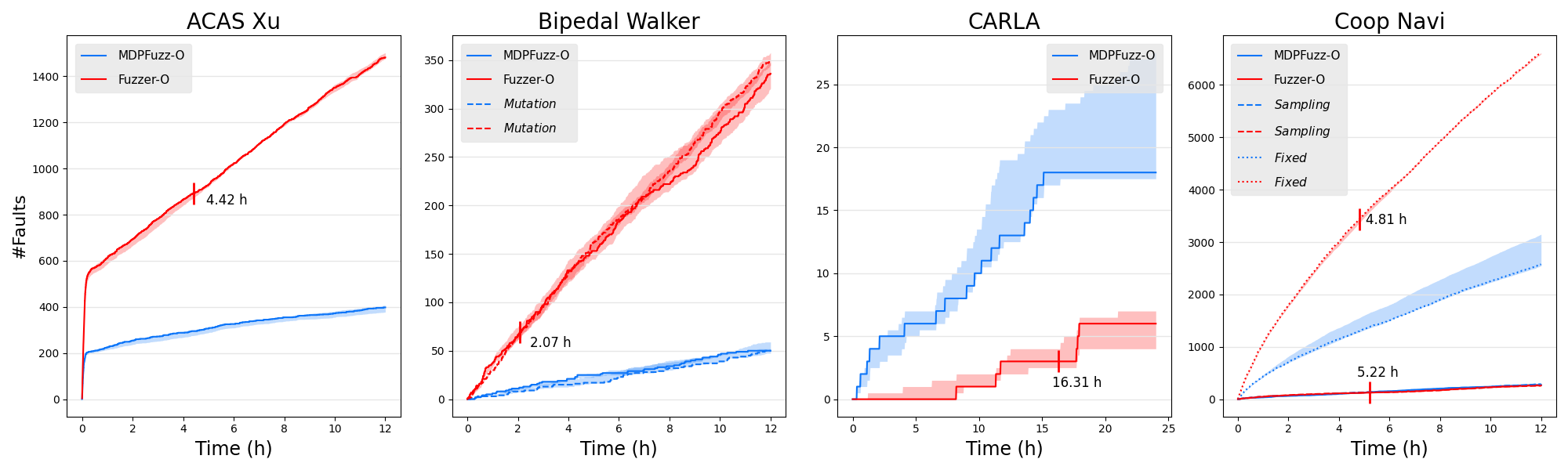}
    \caption{
    Summary of the reproduction results as the number of faults found over time.
    The lines show the median results over 3 executions, and the shaded areas draw the interquartile ranges (IQRs).
    The results of the threats to validity studied are plotted with dashed and dotted lines. 
    The timestamps indicate the time by which Fuzzer-O tests as many states as MDPFuzz-O.
    }
    \label{fig:repro}
    \Description[Summary of the reproduction results as the number of faults found over time.]{}
\end{figure*}

Figure~\ref{fig:repro} reports on the number of faults found during fuzzing, in a similar fashion as the original publication.
To improve readability, we show the results of each use case in separate plots.

\paragraph{Results}

We find consistent results to the original evaluation for CARLA (third column of Figure~\ref{fig:repro}) as well as Coop Navi when the latter is not entirely fixed (last column, bold and dashed lines).
To that regard, we note that the number of test cases executed for the ``fixed'' version (dotted lines, same column) is less than half of the unfixed ones.
This was expected since these flawed versions feed the fuzzers with terminal states.
For all the other fixed use cases except CARLA, we observe that Fuzzer-O outperforms MDPFuzz-O, which is not the case in the original study.
In particular, the long time spent in computing coverage prevents MDPFuzz-O from testing as many inputs as the Fuzzer-O.
Precisely, as shown in Figure~\ref{fig:repro}, for almost all the cases studied, Fuzzer-O needs less than 6 hours to perform the same number of test cases as MDPFuzz-O; the only exception being CARLA, for which it needs 16 hours (of 24 hours, i.e., two thirds of the testing time). 

Furthermore, the GMMs only guide the search towards faults for CARLA; otherwise Fuzzer-O discovers more failures throughout testing.
This can be seen in Figure~\ref{fig:repro}, where the red lines are consistently above the blue ones.
In Bipedal Walker, fixing the mutation operation does not impact the results.
% In ACAS Xu, the results of what we suspected to be the original implementation of the Fuzzer don't align with the original, indicating that the original work evaluates the latter with a version of the code not provided.

\paragraph{Analysis}

Even though MDPFuzz finds faults in the models under test, its lighter counterpart Fuzzer outperforms the former for 3 of the 4 (fixed) use cases studied.
If MDPFuzz's performance for CARLA is consistent with the original results found by~\citeauthor{10.1145/3533767.3534388}, sharing the same conclusions for the two unfixed versions of Coop Navi indicates that the flaws were indeed present when~\citeauthor{10.1145/3533767.3534388} evaluated their proposal\footnote{One could have hoped the bugs to be introduced after the evaluation, as the authors acknowledged to have changed the code thereafter.}.
Our results show that Fuzzer is more applicable and more efficient than MDPFuzz (by avoiding computation coverage).
More importantly, we find that MDPFuzz's additional coverage guidance does not in general favor fault discovery.
Indeed, for all the experiments except CARLA, the baseline detects more faults as soon as fuzzing starts.

We identify several reasons why we do not observe the same results between the two fuzzers as the original evaluation.
First, note that different absolute numbers of faults are not abnormal, since ~\citeauthor{10.1145/3533767.3534388} define the testing budget in time, which makes experiments hardware dependent.
Then, we suppose that the Fuzzer executes significantly more test cases since it does not compute coverages.
Still, we recognize that it does not explain why we find MDPFuzz outperforming its unguided counterpart only for CARLA.
Our hypothesis is that the authors evaluated the Fuzzer by executing the code of MDPFuzz with the line to maintain the pool with low-covered inputs put in comments (as they suggested during our email conversation~\cite{emailconversation}).
% Unfortunately, as mentioned in our code review (Subsection~\ref{subsec:review}), that is not the case in the only implementation of the Fuzzer included in the code, which uselessly computes coverage.

\paragraph{Conclusion}

We confirm that MDPFuzz can detect faults in the decision models tested, with and without coverage guidance.
% However, we don't find that the coverage guidance leads to more faults.
However, in general we don't find that its additional coverage guidance leads to more faults.
This observation is our main motivation to further our investigations and replicate the work. 
Our second motivation is the overall poor quality of~\citeauthor{10.1145/3533767.3534388}' implementation.
Beyond the bugs spotted and the lack of Fuzzer's implementation, the code design can be improved with more modularity and usability.
Indeed, the use cases embed their fuzzer, instead of relying on a single implementation.
As such, it's not easy to adapt it to new use cases, limiting the adoption by researchers and practitioners and making comparisons to and extensions of MDPFuzz difficult.
% ; something that does not align with the general applicability of this policy testing model.
% Lastly, we stress that the current state of the code is alarming.
% For instance, it includes several unused variables, commented instructions and flaws.

\begin{summarybox}{Summary of RQ1 (Reproduction)}{
% We identified several bugs in the original implementation that made us question the validity of the original results.

We identified mismatches between the original implementation and the algorithmic description. 
The provided MDPFuzz and the Fuzzer both find faults in the decision models under test.
However, Fuzzer is significantly better than~\citeauthor{10.1145/3533767.3534388}' findings, as it outperforms MDPFuzz in 3 of the 4 cases studied.
Similarly, in general the coverage guidance does not drive testing towards more faults, which are at the odds of the original results.}
\end{summarybox}

\section{Replicability Study}\label{sec:replication}

In this section, we present how we replicate the original work to answer RQ2 and RQ3.
Through the replication process, we aim to clarify the doubts cast by the reproduction study, to gain insights into the applicability of MDPFuzz and to provide the policy testing research community with usable, modular artifacts.

\subsection{Experimental Design}

We re-implemented from scratch the algorithm proposed by~\citeauthor{10.1145/3533767.3534388}.
Note that we rigorously followed the pseudo-code specified in the paper and did not introduce any adaptations as done by the authors (cf. Section~\ref{subsec:changes}).
We conducted experiments with the corrected versions of the original use cases and three new decision tasks (detailed below).
We compared our replicate of MDPFuzz and Fuzzer (namely, ``MDPFuzz-R'' and ``Fuzzer-R'') with Random Testing (RT), which acts as a baseline and allows us to assess the difficulty of the testing tasks. 
% To that regard, we remind the readers that in the following we refer to these re-implementations as ``MDPFuzz-R'' and ``Fuzzer-R''. 
We measured the number of faults found during testing (as in Section~\ref{sec:reproduction}) and analyzed the time distribution of the three methods.

% expectations of the results
% Based on our previous observations, we expected Fuzzer-R to be competitive with MDPFuzz-R, and our parameter analysis of MDPFuzz-R (RQ2) will not be conclusive.
% Indeed, as explained by the authors (see Subsection~\ref{subsec:changes}), state coverage estimates correlations and, similarly, GMMs are not powerful enough to handle long sequences. 

% \begin{itemize}
%     \item \textit{Reimplementation} We implemented from scratch the algorithm proposed by~\citeauthor{10.1145/3533767.3534388}. We strictly followed the latter and did not implement the changes done by the authors.
%     \item \textit{Original threats to validity} We fixed the issues found in the original implementation (e.g., the use of references in the Coop Navi experiments).
%     \item \textit{Additional use-cases} Replicability studies are also supposed to extend the scope of the evaluation. We therefore added three use-cases: a toy example, Lunar Lander and Cart Pole. This lets us assess the limit of the applicability of the original work.
%     \item \textit{Repeated runs} We accounted for the randomness effect and ran 10 times all the experiments. All our results are shown as the median values and the first and third quantiles.
% \end{itemize}

\subsubsection{Additional Use Cases}
We extended the original evaluation with three commonly used decision-making environments~\cite{towers_gymnasium_2023} for Reinforcement Learning.
They complement the study with diverse testing conditions, varying in complexity and sequence length, and observation spaces, including dimensions and subset types (i.e., $\mathbb{N}$ or $\mathbb{R}$).

\paragraph{Cart Pole} 
In this control problem, a pole is attached to the top of a cart that can move along a horizontal track.
The goal is to keep the pole balanced by pushing the cart to the left and the right.
The initial states specify cart's and pole's position and velocity.

\paragraph{Lunar Lander}
The decision task consists of safely landing a spacecraft on the Moon.
The starting position of the spacecraft is always at the top centre of the space and, likewise, the landing pad is always at the centre of the ground.
The initial situations differ in the shape of the Moon (surrounding the landing pad) and the initial force applied to the space vehicle.
The policy controls the main and orientation engines of the spacecraft.
Note that, as for Bipedal Walker, we consider deterministic executions and thus fix the random seed of the simulator.
As such, the input space is reduced to the initial force applied to the agent (i.e., the landscape is always the same).

\paragraph{Taxi}
This problem depicts a grid world in which the policy picks up passengers and drop them off at their destinations~\cite{dietterich2000hierarchical}.
As such, an initial situation defines the passenger's position and destination as well as the position of the taxi.
At every step, the decision model has six possible actions: moving the taxi (going north, south, east or west) or interacting with the passenger (pick it up or drop it off). 
We enlarge the original version of the environment with a 18x13 map to disable the possibility to enumerate all MDP's possible states.

\begin{figure*}[t]
    \centering
    \includegraphics[width=\textwidth]{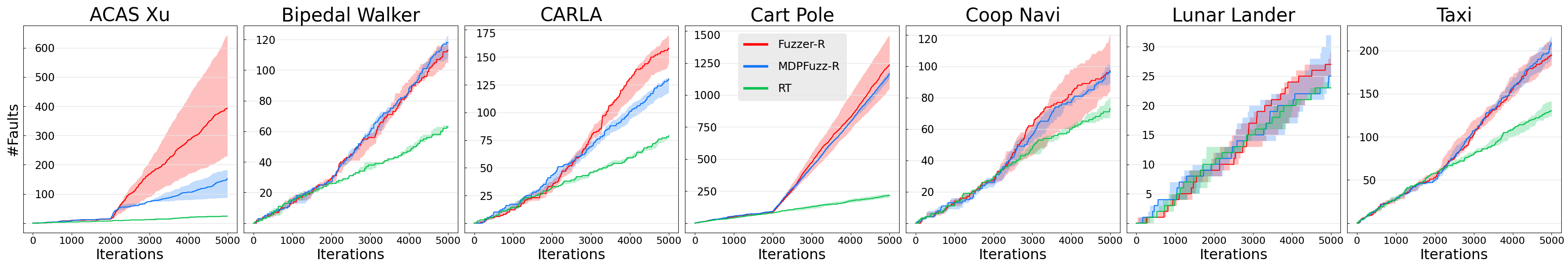}
    \caption{
    Evolution of the number of faults found with the replicate (Fuzzer-R and MDPFuzz-R) and Random Testing (RT). 
    The lines show the median results over 5 executions (3 for CARLA), with their relative IQRs as shaded areas.
    }
    \label{fig:rq1}
    \Description[Evolution of the number of faults found with the replicate (Fuzzer-R and MDPFuzz-R) and Random Testing (RT).]{}{}
\end{figure*}

\begin{figure*}
    \centering
    \includegraphics[width=\textwidth]{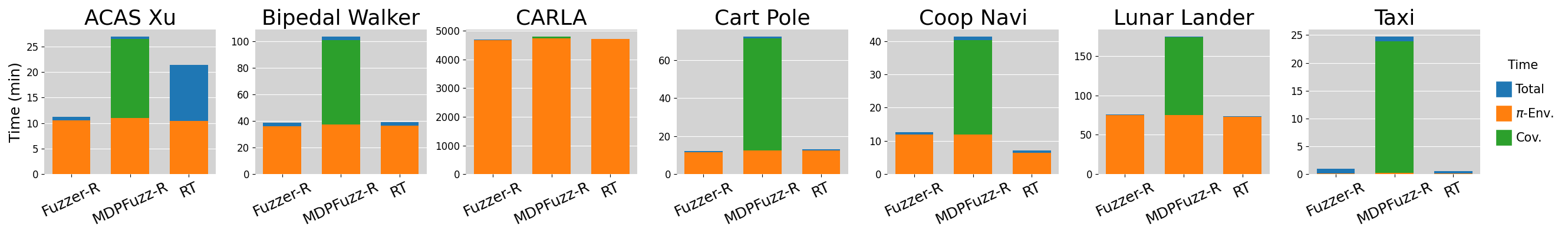}
    \caption{
    Running time of the replication study.
    We report the total, testing ($\pi$-environment interactions) and coverage times of the three methods for each use case.
    The bars show the median values of 5 executions (3 for CARLA), in minutes.
    }
    \label{fig:rq1_time}
    \Description[Running time of the replication study.]{}{}
\end{figure*}

\subsubsection{Mutation}

For Cart Pole and Lunar Lander, we add a small random perturbation to the inputs.
In Taxi, a mutation changes either the position of the car, the position of the passenger, or the destination.

\subsubsection{Test Oracles}
For Cart Pole, the policy fails if the absolute angle of the pole or the position of the cart exceeds threshold values.
In Lunar Lander, a failure occurs if the lander moves outside the viewport or crashes into the ground.
For the Taxi environment, we detect a fault in case of an invalid action (e.g., dropping the passenger while the taxi is still empty) or collision.

\subsubsection{Models Tested}
For Lunar Lander, we test a policy of the freely available Stable-Baselines3 repository~\cite{rl-zoo3} (from which also originates the model tested in Bipedal Walker).
For the Cart Pole and Taxi use cases, we train agents via Q-Learning~\cite{Watkins1992}. 

\subsubsection{Implementation}

We implement Lunar Lander and Taxi in the modified 0.19.0 gym version~\cite{1606.01540} by the original work (containing Bipedal Walker).
We use the Cart Pole version of gymnasium 0.29~\cite{towers_gymnasium_2023} and implement the test executions with Gimitest~\cite{gimitest}, a recent tool for testing RL policies.
We set the maximum number of time steps per simulation $M$ to 400, 1000 and 200 for Cart Pole, Lunar Lander and Taxi, respectively.  
To improve the stability of the executions and to enable reproducibility of the experiments, we fix the random parameters of the underlying MDPs in Lunar Lander and Bipedal Walker.
All the code of the experiments is freely available\footnote{\url{https://github.com/QuentinMaz/MDPFuzz_Replicability_Study_Artifact/tree/master/replication}}.
As for the fuzzers and RT, we have developed them in an independent Python package, also released online\footnote{\url{https://github.com/QuentinMaz/MDPFuzz_Replication}}.
This tool enables the test of any policy with MDPFuzz, Fuzzer or RT for further research.

We ran the three methods with a budget of 5000 tests, and a sampling phase of 1000 iterations for the fuzzers.
By fairness, we deduce from the testing budget of the fuzzers the number of test cases executed during sampling.
We account for randomness effects by repeating all the executions with 5 random seeds (3 for CARLA).
As before, we report median results as well as interquantile ranges.

\subsection{RQ2: Fault Discovery}

We use the same settings as in the reproduction study, i.e. MDPFuzz-R is configured with $K=10$ and $\tau=\gamma=0.01$.

% \begin{table*}[htbp]
%     \centering
%     \input{tables/table}
%     \caption{Time distribution of the three methods for the use-cases studied. 
%     Each row describes the results for case, and methods' results are shown in the columns. We report the median and IQR values of 5 executions (3 for CARLA) in minutes. 
%     Precisely, ``Time'' refers to the total time, ``$\pi$-Env.'' to the total amount of time for executing test cases (agent-environment interactions) and ``Cov.'' to the coverage computation time (only relevant for MDPFuzz, central column).
%     } 
% \end{table*}

\paragraph{Results}

Figure~\ref{fig:rq1} shows the evolution of the faults found during testing.
Except for Lunar Lander, where the three methods perform equally, the fuzzers find more faults than Random Testing.
We observe a significant difference between the fuzzers in ACAS Xu and CARLA.
For ACAS Xu, Fuzzer-R improves on average MDPFuzz-R's results by 175\% at the end of testing.
Likewise, Fuzzer-R finds 33\% more faults than MDPFuzz-R in CARLA. 

The fuzzing techniques also differ in terms of total execution times.
Figure~\ref{fig:rq1_time} shows the running time of the three methods.
For each case study, it details the total, testing ($\pi$-environment interactions) and coverage times (for MDPFuzz-R).
We can see that while Fuzzer-R and RT spend most of their time executing test cases, MDPFuzz-R suffers from an overhead which is significant for all the cases studied except CARLA.
For these cases, the total times increase more than 100\% compared to Fuzzer-R, meaning that MDPFuzz-R actually spends more time computing coverage than testing the models.
Besides, Fuzzer-R seems to be very sensitive to randomness in ACAS Xu, as denoted by the large shaded areas in Figure~\ref{fig:rq1} (first column).
Also, for this use case we measure a substantial overhead for Random Testing.
We looked closely at the data and found out that the latter was caused by the input generation function (borrowed from the original implementation), which requires more time than mutating inputs.

\paragraph{Analysis}

The results of our replicate confirm the ability of the fuzzers to find faults.
Indeed, except for Lunar Lander, the fuzzers beat the random testing baseline, highlighting thus their relevance as dedicated techniques for testing policies.
Likewise, as found previously, we observe a better efficiency of the Fuzzer compared to MDPFuzz.
This was shown in the reproduction study with more test cases for the same time; here with lesser time to perform the same number of executions.
Though, we note that the coverage overhead is not always significant, as shown by the CARLA use case, where executing a test case is longer than coveraging its state sequence.
However, if in the reproduction study the coverage overhead was offset by better MDPFuzz's performance (third column in Figure~\ref{fig:repro}), here Fuzzer-R reveals systematically more faults.

\begin{figure}[t]
    \centering
    \includegraphics[width=\columnwidth]{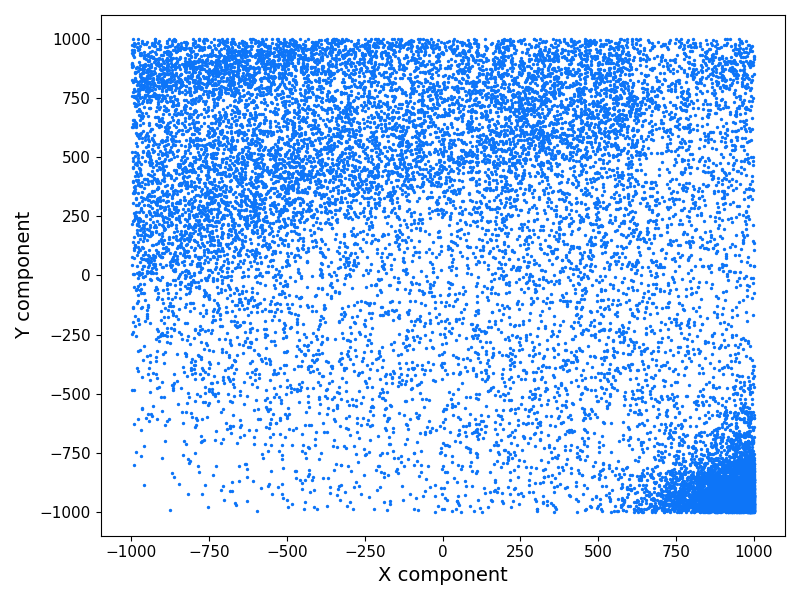}
    \caption{
    Fault distribution of $\pi$ for the Lunar Lander use case, where the inputs describe the initial force applied to the agent.
    The horizontal and vertical axes show the components of the forces in the x and y directions, respectively.
    The input space ($[-1000, 1000]^2$) has been evenly sampled with $4.10^6$ points.
    }
    \label{fig:ll_faults}
    \Description[Fault distribution of $\pi$ for the Lunar Lander use case, where the inputs describe the initial force applied to the agent.]{}{}
\end{figure}

% Lunar Lander investigation?
We further analyze the results for Lunar Lander, since the similarities in the performance of the fuzzers and the random baseline might indicate a possible limitation of the work proposed.
% We tried to find out the reasons behind those results by looking at where the faults are located in the input space.
We investigated the results obtained by closer analysing the fault distribution in the input space.
Our hypothesis was that a draw with RT is likely to occur if the faults are uniformly distributed.
To answer our assumption, we thoroughly test the policy with $4.10^6$ evenly sampled inputs.
Figure~\ref{fig:ll_faults} shows the fault distribution in the input space ($[-1000, 1000]^2$).
We can see that the faults are not uniformly distributed.
Therefore, we would have expected the fuzzers to outperform RT by driving the search towards dense regions (such as the lower right corner in Figure~\ref{fig:ll_faults}).
% The possible explanations can be that (1) the random perturbation of the mutation operator is not high enough or (2) the sampling phase is too short.
% Unfortunately, a deeper investigation is out of the scope of this paper, and we thus deem the draw unanswered. 
While a deeper investigation is out of the scope of this paper, these results reveal a limitation of the mutation-based solution space exploration of the fuzzers.
In any case, we stress that this use case is by far the most difficult testing task, for which $\pi$ has a failure rate of approximately of 0.5\%.

\begin{figure*}[]
    % \centering
    \includegraphics[width=0.75\textwidth,keepaspectratio]{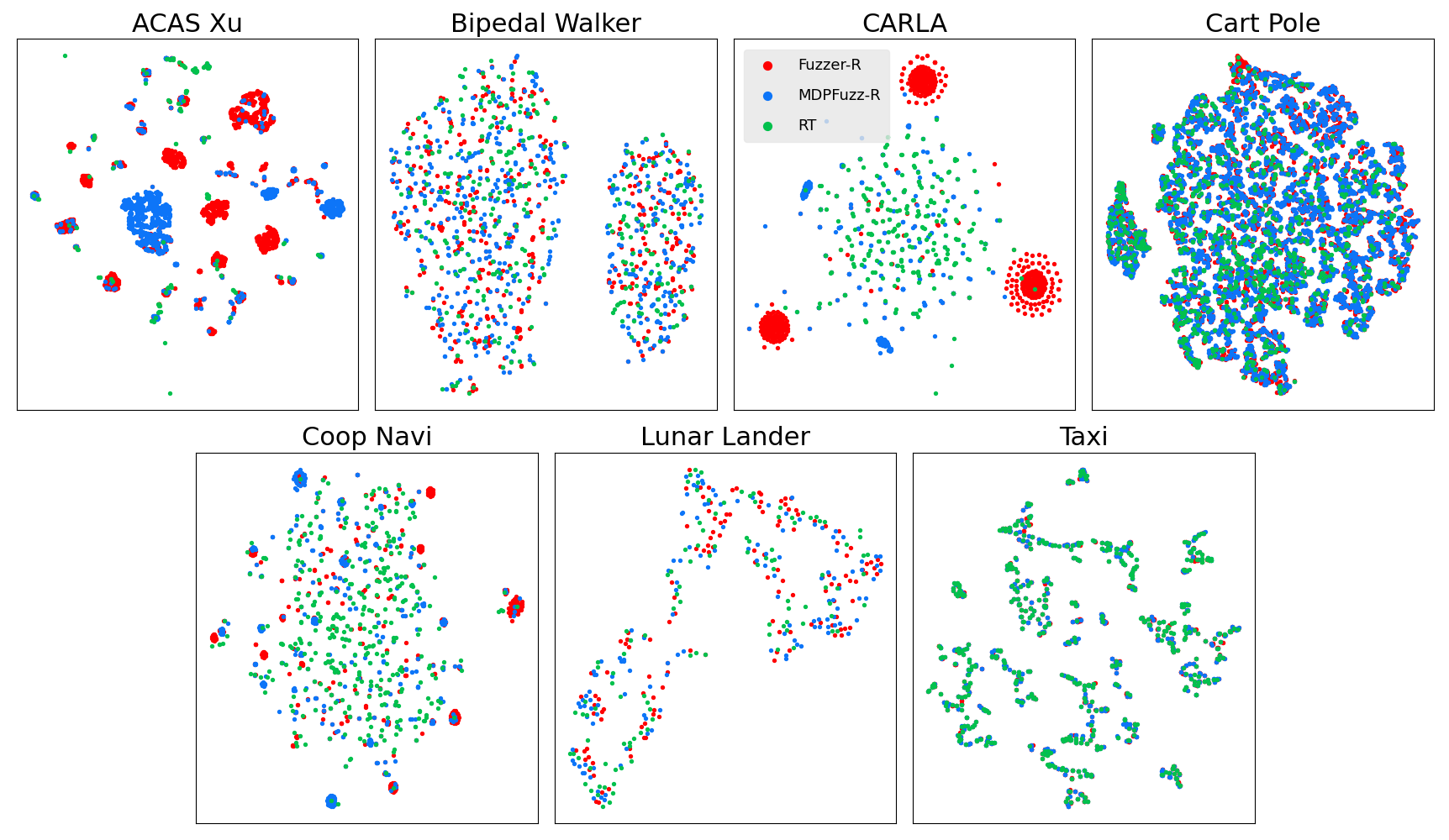}
    \caption{
    TSNE projections of the faults found by the three methods in the replication study (i.e., Fuzzer-R, MDPFuzz-R and RT).
    }
    \label{fig:faults}
\end{figure*}

For sake of completeness, we finally compare the distributions of the faults found by the three methods.
Figure~\ref{fig:faults} shows 2D projections of the faults with TSNE~\cite{Maaten2008VisualizingDU}, as done in the original work.
We can see that the faults are uniformly distributed among the three methods for four of the seven use cases (Bipedal Walker, Cart Pole, Lunar Lander and Taxi).
In CARLA and Coop Navi, the projections indicate that RT finds a wide type of faults (widespread points), while the fuzz-based methods focus on specific clusters less covered by RT.
Eventually, in ACAS Xu each approach seems to reveal distinct faults, even though some of the clusters are shared among the methods (see the first plot in Figure~\ref{fig:faults}).
While these results might exhibit interesting differences in the fault distributions between RT and the fuzzers, we remind that they are 2D projections of high dimensional data points.
If the initial dimensions can be low (2 in Lunar Lander, 4 in ACAS Xu, Cart Pole and Taxi), they increase to 12 and 15 for Coop Navi and Bipedal Walker, respectively. 
To that regard, CARLA is the most extreme case, featuring a 406 dimensional input space.

\begin{summarybox}{Summary of RQ2 (Replication)}{
By outperforming the random testing baseline in 6 of the 7 cases studied, MDPFuzz and Fuzzer show their relevance as testing approaches tailored to policies.
In particular, our replicate further confirms Fuzzer's high efficiency (already found in RQ1), with no significant method overhead.
However, while the reproduction study reveals some usefulness of the coverage guidance (e.g., CARLA), we find that the Fuzzer always finds at least as many failures as MDPFuzz.
As such, we conclude that the coverage guidance proposed by~\citeauthor{10.1145/3533767.3534388}, the key feature of MDPFuzz, is not needed.}
\end{summarybox}

\begin{figure*}[t]
    \centering
    \includegraphics[width=\textwidth]{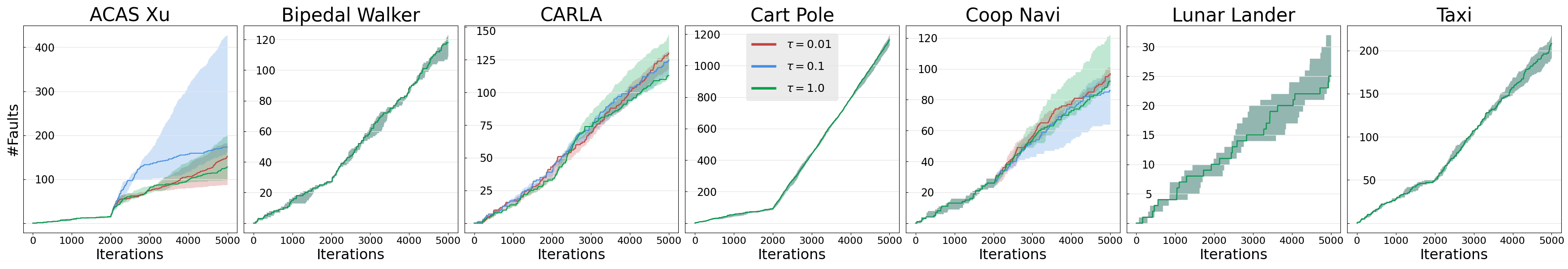}
    \caption{
    Analysis of the parameter $\tau$ on the ability of MDPFuzz-R to find faults in the models under test.
    We use the same settings for $K$ and $\gamma$, and explore $\tau \in [0.01, 0.1, 1.0]$ (red, blue, green). 
    Each plot details the results of a use case.
    The lines are the median values of 5 executions (3 for CARLA) and the shaded areas represent the IQRs.
    }
    \label{fig:rq2_tau}
    \Description[Analysis of the parameter $\tau$ on the ability of MDPFuzz-R to find faults in the models under test.]{}{}
\end{figure*}

\subsection{RQ3: Parameter Analysis}

This last research question aims at investigating the impact of the parameters $K$, $\tau$ and $\gamma$ on MDPFuzz.
In particular, we wonder about its potential sensitivity and whether we can identify a fault-driving configuration which would justify coverage guidance.
Besides, this study will reduce biases that might stem from parameter selection in the previous experiments.

Unfortunately, there is no straightforward way to define them.
We thus explore different values around the ones used so far.
To make the study tractable, we first isolate the parameter $\tau$, since we suspect little to no impact.
Indeed, we observed that most coverage values are either close to zero or infinity, but not spread over a wide range that would be strongly affected by a threshold value.

\subsubsection{Impact of \text{\large{$\tau$}}}

Figure~\ref{fig:rq2_tau} shows the performance of our replicate (i.e., MDPFuzz-R) for three different $\tau$ values: $[0.01, 0.1, 1.0]$.
Even though we observe some differences for ACAS Xu (first column), they are not substantial enough to be more deeply explored.
Besides, we note that we do not aim at fine-tuning the parameters but, rather, at evaluating their effects.
Therefore, we conclude that the exact value of $\tau$ is not critical, as long as the latter distinguishes high coverage values from small ones.
We thus decided to keep the same coverage threshold as before, i.e. $\tau=0.01$.

\subsubsection{Impact of $K$ and \text{\large{$\gamma$}}}

\begin{figure*}[t]
    \centering
    \includegraphics[width=\textwidth]{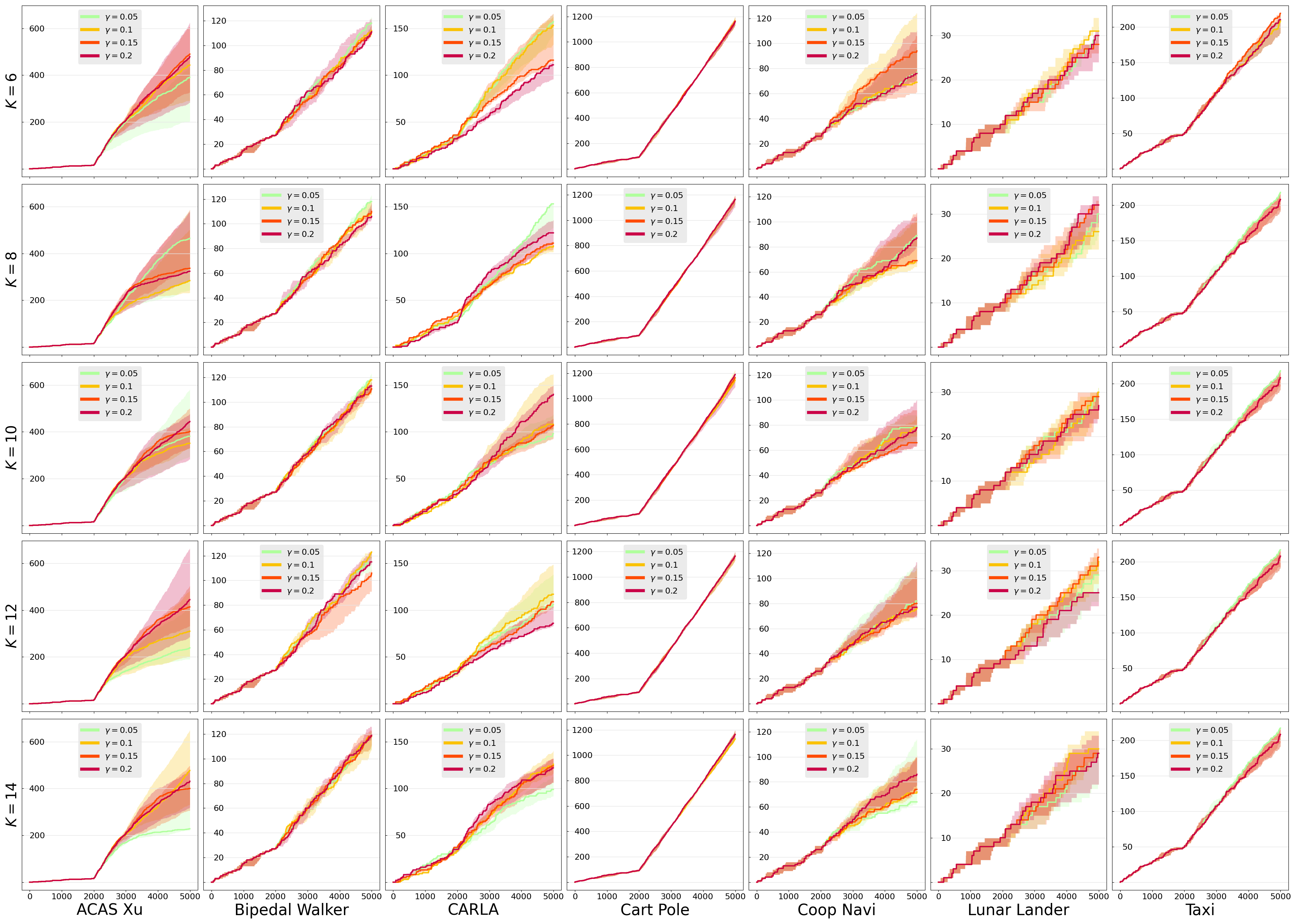}
    \caption{
    Parameter analysis of MDPFuzz with our replicate (i.e., MDPFuzz-R).
    The cases studied span over the columns, while the rows correspond to different $K$ values ($[6, 8, 10, 12, 14]$). 
    Then, each plot details the effect of the parameter $\gamma$ ($[0.05, 0.1, 0.015, 0.2]$).
    We report the median values of 5 executions (3 for CARLA) and show their IQRs as shaded areas.
    }
    \label{fig:rq2}
    \Description[Parameter analysis of MDPFuzz with our replicate (i.e., MDPFuzz-R).]{}{}
\end{figure*}

We investigate a total 20 $(K,\gamma)$ configurations, with $K \in [6, 8, 10, 12, 14]$ and $\gamma \in [0.05, 0.1, 0.015, 0.2]$.
Figure~\ref{fig:rq2} summarizes our findings, where each column shows the results for a use case.
Then, the rows describe different $K$ values, and every curve of a plot corresponds to the median number of faults found with a particular $\gamma$. 

\paragraph{Results}
We do not observe a best single configuration.
In fact, in most cases, we do not find significant differences between them.
Now, if we go into details, we see that $(K,\gamma)$ pairs of smaller values seem to marginally increase the performance (e.g., \text{$8$-$0.05$} for ACAS Xu, \text{$6$-$0.05$} for CARLA and Coop Navi).
Yet, these improvements depend on the case studied.
% and are not meaningful enough to legitimate the use of GMMs (and DynEM) to guide the fuzz framework.

\begin{summarybox}{Summary of RQ3 (Parameter analysis)}{
Our analysis did not reveal significant sensitivity of MDPFuzz to its parameters.
Similarly, we did not find a best single configuration for MDPFuzz that would justify its use of GMMs (and DynEM) as a guidance mechanism. 
We thus question the need of guiding the fuzz framework proposed by~\citeauthor{10.1145/3533767.3534388}, despite their original findings.}
\end{summarybox}

% \section{Discussion}\label{sec:threats}

% In the following, we summarize the key implications of this paper, and discuss the threats of our methodology and experimental evaluation.

% \paragraph{Implications}
% The first key implication results from the differences between our and the original conclusions regarding the relevance of coverage guidance.
% While~\citeauthor{10.1145/3533767.3534388} find that the latter helps detecting faults in the models, we find that it only increases the testing cost without significantly improving the results.
% As such, our investigations highlight the needs to compare and reproduce the peers' work.
% As for the consequences for the policy testing community, the main takeaway is that we do not recommend the coverage model of MDPFuzz.
% Nevertheless, further research is needed to generalize this conclusion to coverage guidance for fuzz-based frameworks.
% Our results also show that simpler approaches with little computation overhead like Fuzzer are enough for fault discovery.

\section{Threats to Validity}\label{sec:threats}

In the following, we discuss the threats of our methodology and experimental evaluation.

\paragraph{Internal Threats} 
We acknowledge that our re-implementation might be biased by the reproduction study.
We decided to take that risk because we wanted to include the original use cases in our evaluation.
As explained in Subsection~\ref{subsec:review}, the implementation of the latter and MDPFuzz are deeply intricate (especially for CARLA), and we had thus to review the code to set apart the testing techniques from the use cases.
Fortunately, this deep investigation let us identify bugs as well as changes in MDPFuzz (not mentioned in the original work).
We mitigate the bias of this prior knowledge by strictly following the algorithms of the original paper and its supplemental material~\cite{10.1145/3533767.3534388}, without making the aforementioned changes (detailed in Subsection~\ref{subsec:changes}).

\paragraph{External threats}
As most empirical works, our results are inherently bound to the cases studied.
We mitigate the possible biases by evaluating our replicate with the original and new use cases.
To that regard, we consider testing tasks of diverse nature (planning and system control), complexity and settings (observation spaces $S$ and sequence lengths $M$).
Moreover, we carefully include in our experiments a random testing baseline, to ensure the relevance of the policy testing methods to begin with.
Similarly, we investigated uncovered configurations for MDPFuzz, to further confirm our conclusions.
Yet, we recognize that one could have selected different parameters, models under test and/or use cases.

\section{Related Work}

In recent years, reproducibility in Machine Learning (ML) has become a major research topic. 
For instance, in her keynote~\cite{icse19keynote} at ICSE 2019, Joelle Pineau invited the Software Engineering community to build more reproducible, reusable, and robust ML-based software systems. 
Likewise,~\citeauthor{henderson2018deep}~\cite{henderson2018deep} listed different research questions and factors, such as hyperparameters and programming languages, that may affect the generalization of results in the field of Reinforcement Learning (RL).
Our work embraces this research requirement by being, to the best of our knowledge, the first replicability study in RL-based policy testing.

In 2023,~\citeauthor{mazouni2023}~\cite{mazouni2023} reviewed the validation and verification techniques for decision-making models, where these authors highlighted the need to explore the limitations of the current results.
Broader studies of ML-based models such as ~\cite{ZHANG2020106296, 10.1007/s10515-022-00337-x, 10.1613/jair.1.12716} revealed the difficulty of guaranteeing reproducible studies in the field of testing learned-based policies for Markov Decision Processes (MDP).

Policy testing has been approached in several ways, like by using genetic algorithms to reveal faults~\cite{zolfagharian2023searchbased}, reinforcement learning~\cite{9712397, 10172658} or search-based techniques~\cite{https://doi.org/10.48550/arxiv.2205.04887}.
Similarly as MDPFuzz,~\citeauthor{Steinmetz_Fišer_Eniser_Ferber_Gros_Heim_Höller_Schuler_Wüstholz_Christakis_Hoffmann_2022}~\cite{Steinmetz_Fišer_Eniser_Ferber_Gros_Heim_Höller_Schuler_Wüstholz_Christakis_Hoffmann_2022} and~\citeauthor{10.1145/3533767.3534392}~\cite{10.1145/3533767.3534392} have proposed to use fuzzing techniques to generate test inputs.
Besides,~\citeauthor{model_based_fuzzer}~\cite{model_based_fuzzer} and~\citeauthor{ast2024}~\cite{ast2024} compare their approach with MDPFuzz. 
It shows that MDPFuzz had a deep impact on the community of scientists who want to find ways to detect faults in complex learned policies for MDP testing. 

Several works have adopted MDPFuzz~\cite{curefuzz, seqdivfuzz} as their main testing approach.
However, none of these approaches relies on the coverage guidance proposed in the original paper.
Precisely,~\citeauthor{seqdivfuzz}~\cite{seqdivfuzz} optimize MDPFuzz's efficiency by aborting the execution of test cases whose state sequence is not diverse enough.
Diversity is inferred at every ``checkpoint'' time step by a sequence diversity model, which is trained before fuzzing.

~\citeauthor{curefuzz}~\cite{curefuzz} propose a variant of MDPFuzz called CureFuzz, which switches the coverage model with a curiosity score, and balances novelty and fault discovery with a multi-objective input selection (instead of sensitivity).
Anyhow, the analysis of current related work shows that MDPFuzz has played a strong role in guiding research towards interesting results in the field of MDP policy testing. 
It is thus crucial to replicate and reproduce its findings in order to ensure the significance of its results. 

\section{Conclusion}\label{sec:conclusion}

In this paper, we studied a 
% widely adopted and impactful 
black-box fuzz testing framework for models solving MDPs called MDPFuzz~\cite{10.1145/3533767.3534388}, through a two-step methodology.
In the first step, we tried to faithfully reproduce the experiment results described in the paper, by using the available open-source~\citeauthor{10.1145/3533767.3534388}' implementation.
However, we found that (1) the coverage-free version of MDPFuzz, i.e., Fuzzer, was not implemented, and (2) the initialization function input a number of states $N$ instead of a duration of 2-hours, thus encouraging us to adapt the code.
Furthermore, we identified unintentional bugs, different values for the parameters of MDPFuzz (compared to the settings of the original evaluation) as well as significant differences with the MDPFuzz methodology specified in the paper.
We contacted the authors, who kindly replied and provided us with explanations regarding the issues found in the implementation.
We eventually executed both the unfixed and fixed versions of the implementation in an attempt to reproduce the original findings and determined whether the bugs were present when~\citeauthor{10.1145/3533767.3534388} evaluated their method.

On the one hand, we confirmed the ability of MDPFuzz to discover faults in the models under test.
On the other hand, we found that its ablated counterpart, Fuzzer, outperforms MDPFuzz in three of the four use cases studied while being less resource-demanding.
Not only these results are at odds with the original ones, but they also made us question the relevance of coverage-guidance, the key feature of MDPFuzz.

As such, in a second step, we replicate the fuzzers, i.e., implementing the algorithm described in the original paper.
We strengthened the original evaluation with three new use cases and included a random testing baseline.
Furthermore, we conducted a parameter analysis of MDPFuzz, to assess its sensitivity and possibly find out an optimal configuration for fault discovery.
The results of our replication are aligned with the ones of the reproduction study.
Precisely, we found that the Fuzzer systematically outperforms MDPFuzz, both in terms of fault detection and test efficiency (despite the parameter analysis).

Therefore, the main takeaway of this paper is that we do not recommend the coverage model of MDPFuzz.
Yet, further research is needed to generalize this conclusion on coverage guidance for fuzz-based frameworks.
Instead, we recommend addressing fault discovery with simpler approaches like Fuzzer, which have little computation overhead.

The second takeaway of this work is about replicability studies.
Indeed, our investigations highlighted the need to compare and reproduce the peers' work in the field of MDP policy testing.
% To that regard, we provide the policy testing community with re-usable artifacts, to foster the research efforts for fuzz-based testing.
To that regard, we implement re-usable artifacts, to foster the research efforts for fuzz-based testing.
Yet, we emphasize the need to improve the research practices in the policy testing community to enable more reliable findings.
In particular, an important practice to adopt is to detail as precisely and exhaustively as possible the experimental settings; something that could be achieved through detailed supplementary material or the definition of a common setting format in the reproducible artifact.
The latter could be supported by a unified framework for evaluating policy testing methods.
This framework could not only standardize evaluation metrics (e.g., fault discovery and distribution), but it could also implement standard use cases that every future work would include in their empirical evaluation besides their more complex and specific main use cases.
Such a joint framework would thus help researchers demonstrate their works and compare the latter to the state-of-the-art techniques.

% We conclude that fuzzing is a promising approach for policy testing but do not recommend the coverage model proposed by~\citeauthor{10.1145/3533767.3534388}.

\section*{Data-Availability Statement}
The artifact of the two empirical studies conducted in this paper is publicly available on GitHub\footnote{\url{https://github.com/QuentinMaz/MDPFuzz_Replicability_Study_Artifact}} as well as on Zenodo~\cite{mazouni_2024_12668777}.
It includes the data and instructions to reproduce the experiments.
We also release the re-implementation of the three methods as a separate, re-usable package\footnote{\url{https://github.com/QuentinMaz/MDPFuzz_Replication}}.

\section*{Acknowledgements}
This work is funded by the Norwegian Ministry of Education and Research, the Research Council of Norway (project AutoCSP, grant number 324674) and is part of the RIPOST Inria-Simula associate team.
The research presented in this paper has benefited from the Experimental Infrastructure for Exploration of Exascale Computing (eX3), which is financially supported by the Research Council of Norway under contract 270053.

\bibliographystyle{ACM-Reference-Format}
\balance 
\bibliography{sample-base}
\end{document}